# Structure and basic magnetic properties of the honeycomb lattice compounds Na$_2$Co$_2$TeO$_6$ and Na$_3$Co$_2$SbO$_6$

L. Viciu[1], Q. Huang[2], E. Morosan[1], H.W. Zandbergen[3], N. I. Greenbaum[1], T. McQueen[1], and R.J. Cava[1]

[1]Department of Chemistry, Princeton University, Princeton NJ 08544

[2]NIST Center for Neutron Research, NIST, Gaithersburg, MD 20899

[3]National Centre for HREM, Department of Nanoscience, Delft Institute of Technology, Al Delft, The Netherlands

**Abstract**

The synthesis, structure, and basic magnetic properties of Na$_2$Co$_2$TeO$_6$ and Na$_3$Co$_2$SbO$_6$ are reported. The crystal structures were determined by neutron powder diffraction. Na$_2$Co$_2$TeO$_6$ has a two-layer hexagonal structure (space group *P6$_3$22*) while Na$_3$Co$_2$SbO$_6$ has a single-layer monoclinic structure (space group *C2/m*). The Co, Te, and Sb ions are in octahedral coordination, and the edge sharing octahedra form planes interleaved by sodium ions. Both compounds have full ordering of the Co$^{2+}$ and Te$^{6+}$/Sb$^{5+}$ ions in the *ab* plane such that the Co$^{2+}$ ions form a honeycomb array; the stacking of the honeycomb arrays differ in the two compounds. Both Na$_2$Co$_2$TeO$_6$ and Na$_3$Co$_2$SbO$_6$ reveal antiferromagnetic ordering at low temperatures, with a metamagnetic transition displayed by Na$_3$Co$_2$SbO$_6$.

**Introduction**

The study of the magnetic properties of low dimensional systems has played an important role in condensed matter physics.[1] The understanding of both the theoretical models and the concepts of low dimensional magnets gained momentum with the discovery of high $T_c$ superconductivity in cuprates, where the magnetic ions are arranged in square planes.[2] Also of recent interest have been triangle-based geometries: when populated with antiferromagnetically coupled magnetic atoms, frustration of the low temperature magnetic states can occur due to conflicting near neighbor spin interactions.[3] In this category, $Na_xCoO_2$ represents one of the richest systems of current interest.[4-8] The layered structures of the $Na_xCoO_2$ phases consist of planes of edge-sharing cobalt-oxygen octahedra interleaved with Na ions.[9] The Co ions have a triangular planar lattice, considered an important factor in driving the properties in these phases.

One variant of the triangular lattice is the honeycomb geometry, where 1/3 of the sites of a triangular lattice are non-magnetic. Compounds with the magnetic ions in this geometry have manifested interesting properties such as spin-glass behavior,[10] spin-flop transitions[11] and even superconductivity.[12] Here we report the synthesis, structural and magnetic characterization of two Co-based compounds, $Na_2Co_2TeO_6$ and $Na_3Co_2SbO_6$, in which the Co ions are arranged in a honeycomb lattice. $Na_3Co_2SbO_6$ was recently described as being trigonal[13] but no other information related to the synthesis, structural or physical property characterization of this phase is available. The Te-based compound has not been previously reported.

**Experimental**

$Na_2Co_2TeO_6$ and $Na_3Co_2SbO_6$ were prepared by mixing $Na_2CO_3$ (Alfa, 99.5%) and $Co_3O_4$ (Alfa, 99.7%) with $TeO_2$ (JMC, 99.9995%) or $Sb_2O_4$ ($Sb_2O_5$ Alfa, 99.5%, annealed for 12 h at 1000°C in air to prepare $Sb_2O_4$). The Co:(Sb,Te) ratio was stoichiometric, but 20% excess $Na_2CO_3$ was added to compensate for loss due to volatilization. The mixtures were thoroughly ground, pressed into pellets ($\phi$=13 mm) and annealed under flowing nitrogen as follows: 8 days at 800°C were necessary for $Na_2Co_2TeO_6$ with two intermediate grindings - one after 4 days and then another one after 2 days of subsequent annealing; $Na_3Co_2SbO_6$ was obtained after 2 days of thermal treatment at 800°C. In all cases, the furnace temperature was increased by 5°C /min until the desired temperature was reached. Slow cooling to room temperature after reaction completion, 5°C /min, was performed. The compounds are both lightly pastel-colored, indicating that they have large band gaps and no significant electrical conductivity or Co mixed valency.

The purity of the samples was analyzed by powder X-ray diffraction using Cu K$\alpha$ radiation and a diffracted beam monochromator. Neutron diffraction data were collected on $Na_2Co_2TeO_6$ and $Na_3Co_2SbO_6$ at the NIST Center for Neutron Research on the high resolution powder neutron diffractometer (BT1) with monochromatic neutrons of wavelength 1.5403 Å produced by a Cu(311) monochromator. Collimators with horizontal divergences of 15', 20' and 7' of arc were used before and after the monochromator and after the sample, respectively. Data were collected in the 2-theta range of 3° to 168° with a step size of 0.05°. The structural parameters were refined using the program GSAS.[14] The sodium content in each sample was determined by the structure refinement, and was in good agreement with that expected from nominal compositions.

The magnetic susceptibilities were measured with a Quantum Design PPMS system. Zero field cooled (ZFC) and field cooled (FC) magnetization data were taken between 1.8 K and 280 K in an applied field of 1 T. Field dependent magnetization data were recorded at 5 K and 30 K for $Na_2Co_2TeO_6$, and at 5 K, 8 K and 12 K for $Na_3Co_2SbO_6$.

**Results**

The purity of the compounds was confirmed by X-ray diffraction analysis. The powder neutron diffraction pattern of $Na_2Co_2TeO_6$ was indexed with a hexagonal cell in the space group $P6_322$ (No.182), while that of $Na_3Co_2SbO_6$ was indexed with a centered monoclinic cell in the space group $C2/m$ (No. 12). Structural analysis by the Rietveld method based on the powder neutron diffraction data was performed for both compounds. The cell constants, as determined by powder neutron diffraction, are presented in Tables 1 and 2. $Na_2Co_2TeO_6$ displays a primitive, two layer hexagonal structure, while $Na_3Co_2SbO_6$ has a single layer monoclinic structure, very similar to what is seen in $Na_3Cu_2SbO_6$.[13]

The neutron powder diffraction pattern for $Na_2Co_2TeO_6$ shows it to have no observable dimensional deviations from hexagonal symmetry. Its honeycomb-based crystal structure was found to be very well described in space group $P6_322$. Two symmetry-independent Co ions within the honeycomb layer, one type of Te, one type of oxygen, and three independent Na ion positions were found. Refinements were carried out with atoms of the same type constrained to have the same thermal parameters. The Na distribution between the honeycomb layers is highly disordered. The geometry of the oxygen array in the planes neighboring the Na planes creates triangular prismatic sites for the Na, exactly as is found in two-layer $Na_xCoO_2$[9]. In the $Na_xCoO_2$ structures, Na is found to occupy triangular prisms of two types: one type that shares only edges with the metal octahedra in the planes above and below, and one that shares faces with the metal octahedra in the planes above and below. The same situation is found for $Na_2Co_2TeO_6$. In this case, though, there are two types of face sharing possible, as the triangular prisms can share faces with one Co octahedron and one Te octahedron in the planes above and below, or two Co octahedra. The final refinements (Table 1) showed that the Na ions partially occupy all three types of available triangular prismatic sites, in different amounts: 71% of the Na are in the triangular prisms that share edges with octahedra in the honeycomb layers, 23% are in the prisms that share faces with the Co and Te, and 6% are in sites that share faces with two Co. The reason for this distribution is not understood, but as in the case of two-layer $Na_xCoO_2$, the fact that the face sharing positions are occupied at all indicates that the difference in energy that occurs on sharing faces with the octahedra in neighboring planes is not as important as other factors such as Na-Na interactions in determining the Na distribution. Again as in the case of $Na_xCoO_2$,[15,16] some of the Na ions were found to be displaced towards the faces of their triangular prisms. Refinements in which the Na were left in the centers of their coordination polyhedra resulted in worse agreement factors ($\chi^2$=1.47, $wR_P$=6.72, $R_P$=5.26) than for the final model, and physically unrealistic thermal parameters for the Na. It was found that the displacement of Na(2) did not significantly improve the agreement factors or the thermal parameters, so in the final refinement, only Na(1) and Na(3) were allowed to relax from their ideal positions; they were found to displace towards the faces of the prisms. The freely refined occupancies of the three Na sites, which are all partially occupied, yield a formula within experimental uncertainty of the ideal $Na_2Co_2TeO_6$ composition. The final model provides an excellent fit to the data. The observed, calculated, and difference plots for the refinement based on this model are presented in Figure 1. The refined structural parameters are presented in Table 1. The Na distribution is shown in Figure 4.

A classic two-layer structure, similar to that of $Na_{0.7}CoO_2$,[9] but with a larger in-plane unit cell to accommodate the honeycomb-type cobalt-tellurium ordering is found for $Na_2Co_2TeO_6$. No Co-Te intermixing was found for the metal atom sites: the atoms are fully ordered. The crystal structure is presented in Figure 3 a and b. Selected bond distances are shown in Table 3. The two independent Co sites have very similar Co-O bond distances within their octahedra, 2.11 Å and 2.13 Å for Co(1) and Co(2), respectively. The $Te^{6+}O_6$ octahedron is significantly smaller (Te-O bond length = 1.95 Å), as expected. The bond valence sums are consistent with the assignment of $Co^{2+}$ and $Te^{6+}$ as formal oxidation states. The two-layer stacking sequence found for $Na_2Co_2TeO_6$ is distinctly different from that found for the Cu analog, $Na_2Cu_2TeO_6$,[17] which, though monoclinic, is based on a pseudo-hexagonal three-layer stacking sequence.

$Na_3Co_2SbO_6$ was found here to be isostructural with $Na_3Cu_2SbO_6$. It has a single layer structure, well described in the *C2/m* space group. The structure displays one type of Co, one type of Sb, two independent sodium ions and two different oxygen sites. The refinement, using starting positions[9] from $Na_3Cu_2SbO_6$ quickly converged to a final structural model. No mixing was found among the Co and Sb, and the freely refined Na site occupancies were found to be within experimental uncertainty of full occupancy, yielding the ideal composition. All site occupancies were therefore set to 1 in the final refinements. The final structural parameters are presented in Table 2. The observed, calculated, and difference plots for the neutron diffraction data, using the final model, showing the excellent fit of the model to the data, are presented in Figure 2. The crystal structure is presented in Figure 3 a and c. Selected bond distances are shown in Table 3. The monoclinic symmetry allows for the possibility of distortion of the $Co^{2+}$-O octahedron, but the six Co-O distances are in the range of 2.12-2.13 Å, showing that no substantial distortion has occurred. The Co-O bond lengths are very similar to those observed for $Na_2Co_2TeO_6$ (2.11-2.13 Å). The $SbO_6$ octahedron in $Na_3Co_2SbO_6$ has a size ($Sb^{5+}$-O distances 2.03-2.06 Å) very similar to that found for the Co-O octahedron. The bond valence sums are consistent with the assignment of $Co^{2+}$ and $Sb^{5+}$ as formal oxidation states.

The temperature- and field-dependent magnetization data for $Na_2Co_2TeO_6$ and $Na_3Co_2SbO_6$ are shown in Figures 5-8. Both compounds display Curie-Weiss behavior above approximately 150 K. Deviations from the high temperature behavior at lower temperatures suggest the presence of either low dimensional ordering or short range magnetic fluctuations (Figures 5 and 6). The magnetic moments inferred from the linear fits of the inverse susceptibility data are $\mu_{eff}$ = 4.97 $\mu_B$ and $\mu_{eff}$ = 4.90 $\mu_B$ per Co for $Na_2Co_2TeO_6$ and $Na_3Co_2SbO_6$ respectively. These moments are in agreement with the commonly observed value of 4.8 $\mu_B$ for $Co^{2+}$ ions in a high spin $d^7$ configuration.[18]

The temperature dependent magnetization data for $Na_2Co_2TeO_6$ show a maximum near 20 K (top inset in Figure 5), characteristic of a magnetic transition that is predominantly antiferromagnetic (AFM) in character. From the maximum in d(M*T)/dT (lower inset in Figure 5), the Néel temperature $T_N$ is determined to be 18 K. A second transition near 9 K is also apparent in the derivative data, which may be associated with spin reorientation. Both transition temperatures are marked by vertical arrows in the figure. Although the positive Weiss temperature $\theta_W$ ($\theta_W = 7$ K) suggests the presence of weak ferromagnetic (FM) interactions, the magnetization versus magnetic field (M vs. H) data recorded at 5 K and 30 K (Figure 6) give no clear evidence of a hysteresis loop. The curve at 5 K has an upward curvature, however, confirmed by its derivative (inset to Figure 6), suggesting the presence of a broad metamagnetic transition. The fact that this upward curvature is not present in the 30 K data indicates that at this temperature the system is in the paramagnetic state, consistent with the temperature-dependent data and the $T_N$ value of 18 K.

The temperature- and field-dependent magnetization data for $Na_3Co_2SbO_6$ are presented in Figures 6 and 8. The ZFC/FC susceptibility data in an applied magnetic field H = 1 T are indistinguishable and display a broad maximum below 5 K (top inset in Figure 6). This again is indicative of a primarily AFM transition, with the ordering temperature determined from the maximum in d(M*T)/dT to be $T_N = 4.4$ K, despite a positive Weiss temperature $\theta_W = 15.6$ K. The M(H) data (Figure 8) show no hysteresis, but do clearly indicate the existence of a metamagnetic transition that broadens and moves to lower fields as T increases (inset to Figure 8). The critical field values for the transition are seen in the maxima in the dM/dH plots shown in the inset. A field of only 1 T is required to make the Co spins reorient at T = 5 K.

**Discussion and Conclusions**

The crystal structures of $Na_2Co_2TeO_6$ and $Na_3Co_2SbO_6$ consist of planes of edge-sharing octahedra of $CoO_6$ mixed with $TeO_6$ and $SbO_6$ octahedra, respectively, in an ordered fashion: the $Co^{2+}$ ions form a honeycomb lattice, where each Co has three Co near neighbors, and the non-magnetic $Te^{6+}$ or $Sb^{5+}$ ions sit in the middle of each Co honeycomb. The projection along the $c$ axis of $Na_2Co_2TeO_6$ shown in Figure 3 illustrates the type of network formed by Co and Te/Sb in the $ab$ plane in both compounds. This ordered distribution of the metallic ions results in an enlargement of the in-plane hexagonal cell dimension of $Na_2Co_2TeO_6$ over that of the simple triangular array, $a_{HEX}$, to $a_{HEX} \times \sqrt{3}$. The $ab$ plane is dimensionally hexagonal in $Na_3Co_2SbO_6$ as well, though the overall symmetry of the structure is monoclinic.

In $Na_3Cu_2SbO_6$ and $Na_2Cu_2TeO_6$, due to the Jahn-Teller effect manifested by the Cu ions, the Cu honeycomb lattice is quite distorted, and can be considered a lattice of Cu dimers.[13,15] No such distortion is found in the Co honeycomb network for either $Na_2Co_2TeO_6$ or $Na_3Co_2SbO_6$, though in the latter case it would be allowed by symmetry. Slightly smaller honeycombs are found in $Na_2Co_2TeO_6$ than in $Na_3Co_2SbO_6$ due to the smaller size of the $Te^{6+}$ ion.

The sodium distribution found in $Na_2Co_2TeO_6$ is different from that in $Na_3Co_2SbO_6$. In $Na_2Co_2TeO_6$, a disordered Na lattice arrangement is seen, whereas in $Na_3Co_2SbO_6$ the Na ions are ordered in a simple, completely filled triangular array. Although the average structure shows the Na atoms to partially occupy sites between the layers in a disordered fashion in $Na_2Co_2TeO_6$, short range ordering may be present.

The most important difference between the two structures is the stacking of the honeycomb arrays. In the case of $Na_2Co_2TeO_6$, the Co honeycomb arrays are staggered (Figure 9a) while in $Na_3Co_2SbO_6$ the Co honeycomb arrays stacked directly on top of each other (Figure 9b) similar to the Cu honeycomb planes in $Na_3Cu_2SbO_6$.[13] The difference in stacking of the honeycomb layers is expected to strongly impact the magnetic interactions.

The magnetic susceptibility data for both compounds are consistent with the presence of $Co^{2+}$ ions in the high spin configuration (S=3/2). The in-plane Co-Co distances within the honeycomb for the two compounds are similar (3.052 Å for $Na_2Co_2TeO_6$ and 3.098 Å for $Na_3Co_2SbO_6$), and the superexchange pathways, through ~90 degree Co-O-Co bonds between edge-sharing octahedra, are also very similar. Thus, it is not surprising that the Curie Weiss thetas are ferromagnetic in nature and similar in magnitude (7K for $Na_2Co_2TeO_6$ and 15.5K for $Na_3Co_2SbO_6$). The compounds are not isomorphous, however, due to differences in the interplanar stacking. This may be the origin of the different magnetic ordering temperatures (18 K for the Te compound and 4.4 K for the Sb compound). The difference in stacking may also lead to the difference in behavior in an applied magnetic field. Clear evidence for a metamagnetic transition is seen in $Na_3Co_2SbO_6$ at low fields but the situation is less clear in the Te compound, where there is only a broad increase in the magnetization with applied field.

$BaCo_2(AsO_4)_2$ and $BaCo_2(PO_4)_2$ are examples of well studied systems in which the $Co^{2+}$ ions are also located on a honeycomb lattice.[1,19] These systems represent classical examples of 2D XY magnetic models, and have different magnetic ground states. Some of the magnetic properties observed for the current compounds, i.e. the metamagnetism, are reminiscent of what is seen in those materials. A detailed study of the magnetic behavior of single crystals of $Na_2Co_2TeO_6$ and $Na_3Co_2SbO_6$ would be of particular interest because their honeycomb lattices are similar in size and geometry but differ in stacking. Such a comparison might illuminate the effects of interlayer interactions on the low temperature magnetic state of honeycomb lattices.

**Acknowledgements**

This work was supported by the US department of Energy, Division of Basic Energy Sciences, grant DE-FG02-98-ER45706. T. McQueen gratefully acknowledges support by the national science foundation graduate research fellowship program. Certain commercial materials and equipment are identified in this report to describe the subject adequately. Such identification does not imply recommendation or endorsement by the NIST, nor does it imply that materials and the equipment identified is necessarily the best available for the purpose.

**Table 1.** Crystallographic data for $Na_2Co_2TeO_6$ in the space group $P\,6_3\,2\,2$ (No. 182), $a$=5.2889(1)Å, $c$=11.2149(4)Å, Volume = 271.68(1) Å$^3$.

| Atom | Wyckoff position | x | y | z | $Uiso$*100 | $Occ$ |
|---|---|---|---|---|---|---|
| Co(1) | 2b | 0 | 0 | ¼ | 0.8(1) | 1 |
| Co(2) | 2d | 2/3 | 1/3 | ¼ | 0.8(1) | 1 |
| Te | 2c | 1/3 | 2/3 | ¼ | 0.9(1) | 1 |
| O | 12i | 0.6432(6) | -0.0252(4) | 0.3452(1) | 1.50(4) | 1 |
| Na(1) | 12i | 0.24(1) | 0.65(2) | -0.026(3) | 0.8(3) | 0.081(7) |
| Na(2) | 2a | 0 | 0 | 0 | 0.8(3) | 0.11(3) |
| Na(3) | 12i | 0.702(3) | 0.066(2) | -0.006(2) | 0.8(3) | 0.234(6) |

$\chi^2$=1.39; wRp=6.53; Rp=5.05

**Table 2.** Crystallographic data for $Na_3Co_2SbO_6$ in the space group *C 2/m* (No. 12), *a*=5.3681(2)Å, *b*=9.2849(4)Å, *c*=5.6537(2)Å, β=108.506(4)°, Volume= 267.22(2) Å$^3$.

| Atom | Wyckoff position | x | y | z | *Uiso*\*100 | *Occ* |
|---|---|---|---|---|---|---|
| Co | 4g | 0 | 2/3 | 0 | 0.23(12) | 1 |
| Sb | 2a | 0 | 0 | 0 | 1.86(17) | 1 |
| O(1) | 8j | 0.2744(8) | 0.3410(3) | 0.7971(5) | 1.25(2) | 1 |
| O(2) | 4i | 0.2463(10) | 0.5 | 0.2056(10) | 1.25(2) | 1 |
| Na(1) | 2d | 0 | 0.5 | 0.5 | 1.58(5) | 1 |
| Na(2) | 4h | 0.5 | 0.3296(9) | 0.5 | 1.58(5) | 1 |

$\chi^2$=1.01; wRp=5.31%; Rp=4.31%.

**Table 3.** Selected bond distances for $Na_2Co_2TeO_6$ and $Na_3Co_2SbO_6$ at room temperature. BVS= bond valence sum.

| $Na_2Co_2TeO_6$ | | $Na_3Co_2SbO_6$ | |
|---|---|---|---|
| Co(1)-O×6 | 2.112(3) | Co-O(1)×2 | 2.132(4) |
| Co(2)-O×6 | 2.131(3) | Co-O(2)×2 | 2.124(3) |
|  |  | Co-O(1)×2 | 2.125(4) |
| BVS | 1.93 (Co(1)) |  | 1.85 |
|  | 1.83 (Co(2)) |  |  |
| Te-O ×6 | 1.951(2) | Sb-O(1)×4 | 2.026(3) |
|  |  | Sb-O(2)×2 | 2.057(6) |
| BVS | 5.98 |  | 5.10 |
| Na(1)-O* | 2.50(5) | Na(1)-O(1)×4 | 2.367(4) |
| Na(2)-O | 2.515(3) | Na(1)-O(2)×2 | 2.427(6) |
| Na(3)-O* | 2.49(1) | Na(2)-O(1)×2 | 2.368(4) |
|  |  | Na(2)-O(1)×2 | 2.431(6) |
|  |  | Na(2)-O(2)×2 | 2.385(6) |
| Co-Co | 3.054 ×3 |  | 3.095 ×1 |
|  |  |  | 3.098 ×2 |

* The bond length is averaged over 6 different bonds.

**References**

1. "Magnetic properties of layered transition metal compounds" edited by L. J. de Jongh, Kluwer Academic Publisher, 1990.

2. J. G. Bednorz, K.A Muller, *Z. Phys. B* **64**, 189 (1986).

3. A.P. Ramirez, *Annual Rev. Mater. Sci.* **24** 453 (1994)

4. M.G.S.R. Thomas, P.G. Bruce, J.B. Goodenough, Solid State Ionics, 17, 13, (1985).

5. I. Terasaki, Y. Sasago, K. Uchinokura, *Phys. Rev. B* **56**, R12685 (1997)

6. Y. Y. Wang, N.S. Rogado, R.J. Cava, N.P. Ong, *Nature*, **423**, 425 (2003)

7. M.L.Foo, Y.Wang, S.Watauchi, H.W. Zandbergen, T. He, R.J. Cava, N.P.Ong, *Phys. Rev. Lett.* **92**, 247001 (2004)

8. K. Takada, H. Sakurai, E. Takayama-Muromachi, F. Izumi, R.A. Dilanian, and T. Sasaki, *Nature*, **422**, 53 (2003)

9. Claude Fouassier, Guy Matejka, Jean-Maurice Reau, Paul Hagenmuller, *J. Solid State Chem.* **6**, 532 (1973).

10. M. Bieringer, J.E. Greedan, G.M. Luke, *Phys. Rev.B* **62**, 6521 (2000)

11. D. J. Goossens, A.J. Studer, S.J.Kennedy, T.J. Hicks, *J. Phys.:Condens. Matter* **12**, 4233 (2000).

12. S. Shamoto, T. Kato, Y. Ono, Y. Miyazaki, K. Ohoyama, M. Ohashi, Y. Yamaguchi, T. Kajitani, *Physica C*, *306*, 7 (1998).

13. O.A. Smirnova, V.B. Nalbandyan, A.A. Petrenko, M. Avdeev, *J. Solid State Chem.* **178**, 1165 (2005).

14. A. Larson, A. and R.B. Von Dreele, GSAS: Generalized Structure Analysis System; Los Alamos National Laboratory: Los Alamos, NM (1994).

15. Q. Huang, M.L.Foo, R.A. Pascal, Jr., J.W.Lynn, B.H.Toby, T. He, H.W. Zandbergen, R.J.Cava, Phys. Rev B, 70, 184110 (2004).

16. J.D. Jorgensen, M. Avdeev, D.G. Hinks, J.C. Burley, and S. Short, *Phys. Rev. B* **68**, 214517 (2003)

17. Jianxiao Xu, Abdeljalil Assoud, Navid Soheilnia, Shahab Derakhshan, Heather L. Cuthbert, John E. Greedan, Mike H. Whangbo, and Holger Kleinke, *Inorg. Chem.*, **44,** 5042, (2005)

18. C. Kittel, *Introduction in Solid State Physics*, John Wiley & Sons, Inc, 5$^{th}$ Ed, 1976.

19. L.P. Regnault, J. Rossat-Mignod, *Physica,* 86-88B, 660, (1977).

**Figure Captions:**

**Figure 1.** Observed (crosses) and calculated (solid line) neutron diffraction intensities for $Na_2Co_2TeO_6$ at 295K. Vertical bars show the Bragg peak positions. Regions in the pattern in the vicinity of two impurity peaks (39.6-40.7 and 59.8-60.1 degrees) were omitted from the refinement. The difference plot is shown at the bottom.

**Figure 2.** Observed (crosses) and calculated (solid line) neutron diffraction intensities for $Na_3Co_2SbO_6$ at 295K. Vertical bars show the Bragg peak positions. The difference plot is shown at the bottom.

**Figure 3.** The crystal structures of $Na_2Co_2TeO_6$ and Na3Co2SbO6 (a) The layer of edge-shared $MO_6$ octahedra, showing the honeycomb Co array. (b) View perpendicular to the honeycomb layers in Na2Co2TeO6 (c) view perpendicular to the honeycomb layers in Na3Co2SbO6. Inside the octahedra, small dark grey spheres represent Co while bigger black spheres represent Te. The sodium ions are represented by light gray spheres. Oxygens are at the vertices of the octahedra.

**Figure 4.** The Na distribution in $Na_2TeCo_2O_6$. All Na positions are partially occupied. The oxygens in the layers above and below are shown as red spheres. They form three types of triangular prismatic sites for Na. These are marked with their triangular bases in the figure: Na(1) shares faces with one Co and one Te octahedron, and contains approximately 23% of the Na; Na(2) shares faces with two Co octahedra, and contains approximately 6% of the Na; and Na(3) shares only edges with the Co and Te octahedra in the layers above and below (approximately 71% of the Na). Na(1) are grey, Na(2) cyan and Na(3) blue spheres, respectively.

**Figure 5.** The temperature dependence, at an applied field of H = 1 T, of the magnetization for $Na_2Co_2TeO_6$. The dashed line shows the fit of the high temperature data to a Curie-Weiss law. The insets are: upper – detail of low temperature molar susceptibility; lower – low temperature d(MT)/dT at H = 1 T. Significant features at 18 and 9 K are seen in the derivative data.

**Figure 6.** The temperature dependence, at an applied field of H = 1 T, of the magnetization for $Na_2Co_2SbO_6$. The dashed line shows the fit of the high temperature data to a Curie-Weiss law. The insets are: upper – low temperature susceptibility at H = 1 T; lower – low temperature d(MT)/dT at H = 1 T: a feature is seen at 4.4K.

**Figure 7.** The field dependence at T = 5 K and 30 K of the magnetization for $Na_2Co_2TeO_6$. The inset shows the magnetization derivative dM/dH vs magnetic field at T = 5 K and 30 K.

**Figure 8.** The field dependence at T = 5 K, 8 K and 12 K, of the magnetization for $Na_2Co_2SbO_6$. The inset shows the magnetization derivative dM/dH at T = 5 K, 8 K and 12 K. A metamagnetic transition is clearly seen.

**Figure 9.** View of adjacent Co-M (M=Te, Sb) planes in $Na_2Co_2TeO_6$ and $Na_3Co_2SbO_6$, showing the stacking of the cobalt honeycomb lattices. In both cases, the planes have been tilted slightly so that all the metal atoms are visible: for the Te compound the metals are directly on top of each other in the direction perpendicular to the plane while in the Sb case they are shifted slightly. For $Na_2Co_2TeO_6$ (a), the stacking of the honeycomb Co lattices in neighboring planes is staggered, while, for (b) $Na_3Co_2SbO_6$, the Co honeycombs stack nearly directly on top of one another. Small dark grey spheres represent Co, bigger black spheres are Sb and big dark gray crossed spheres are Te.

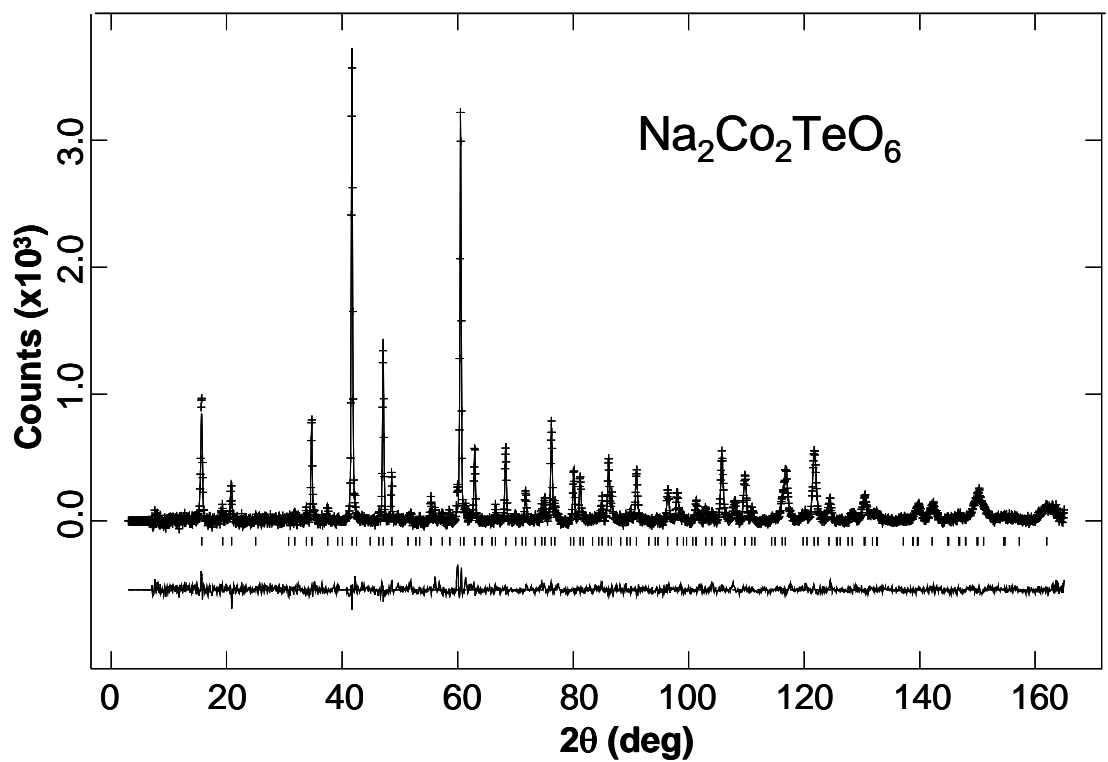

**Figure 1**

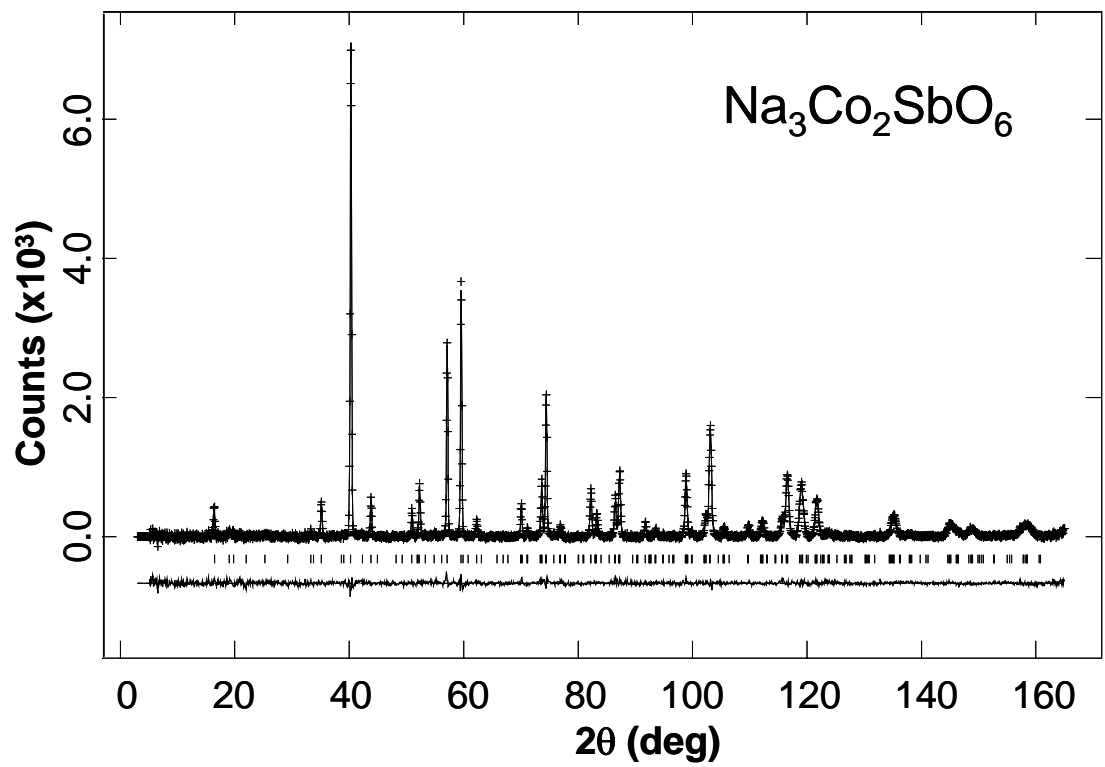

Figure 2

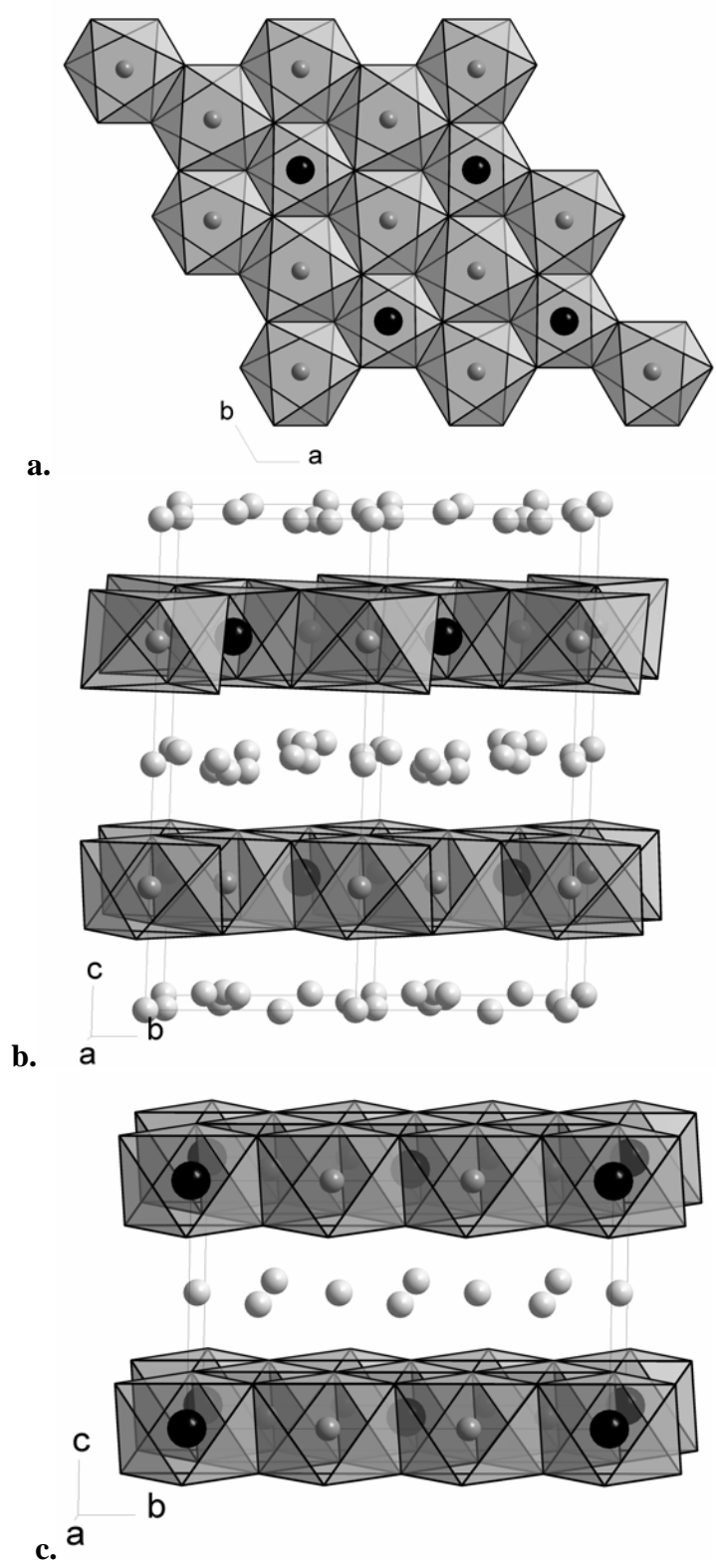

**Figure 3.**

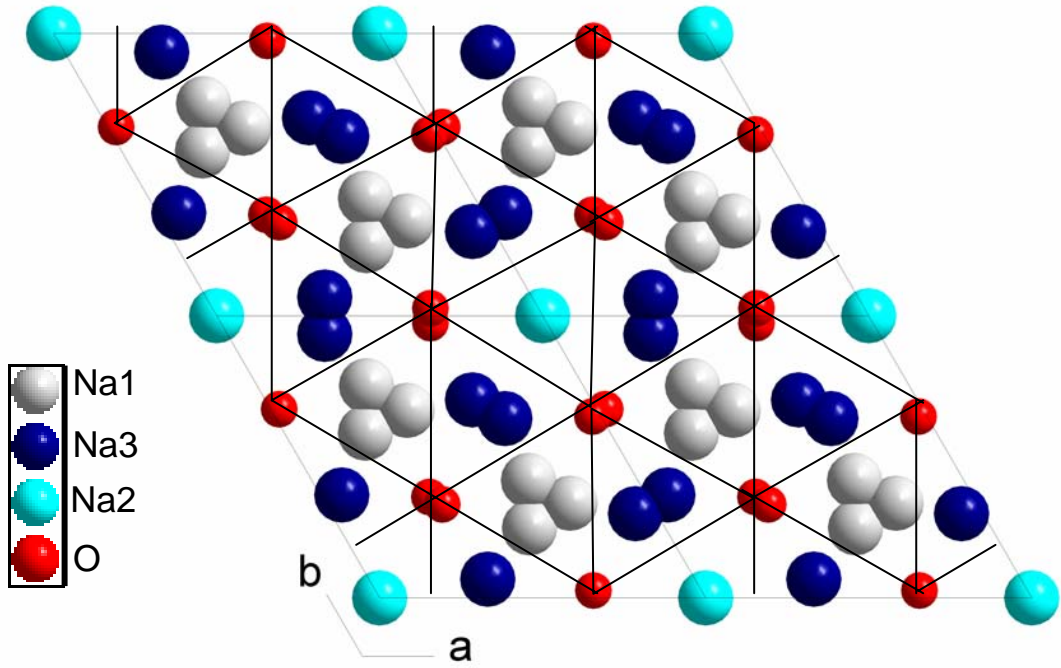

**Figure 4**

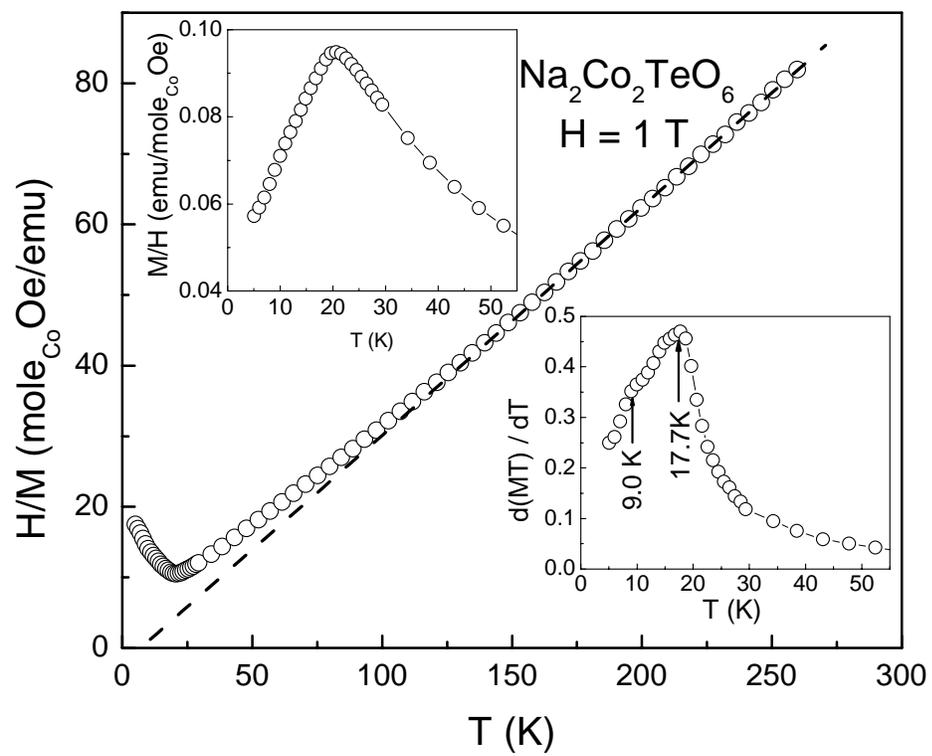

**Figure 5**

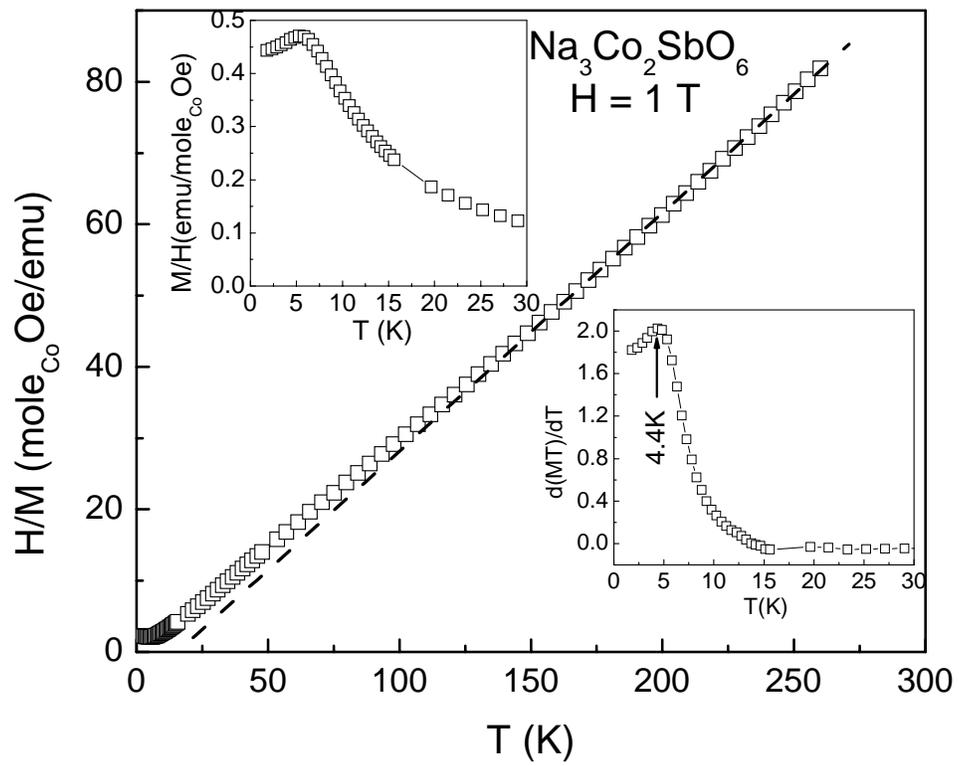

**Figure 6**

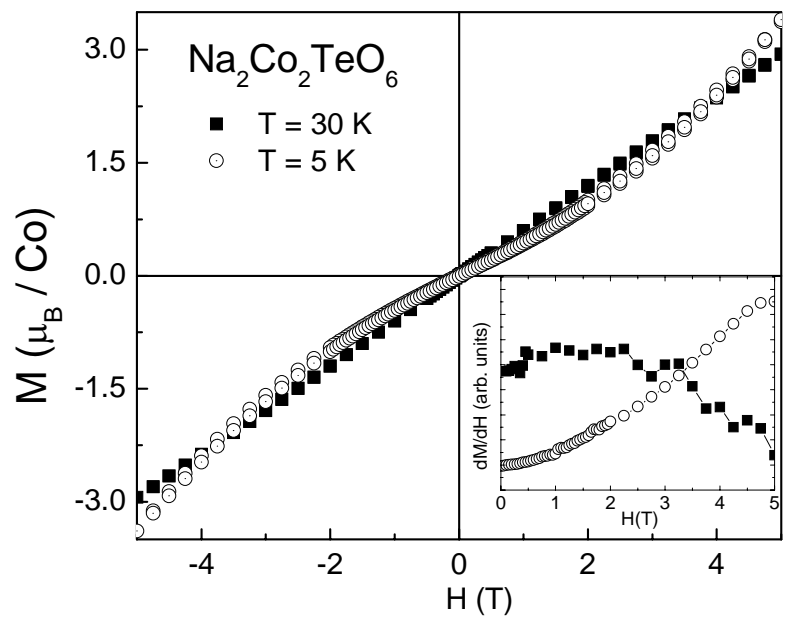

**Figure 7**

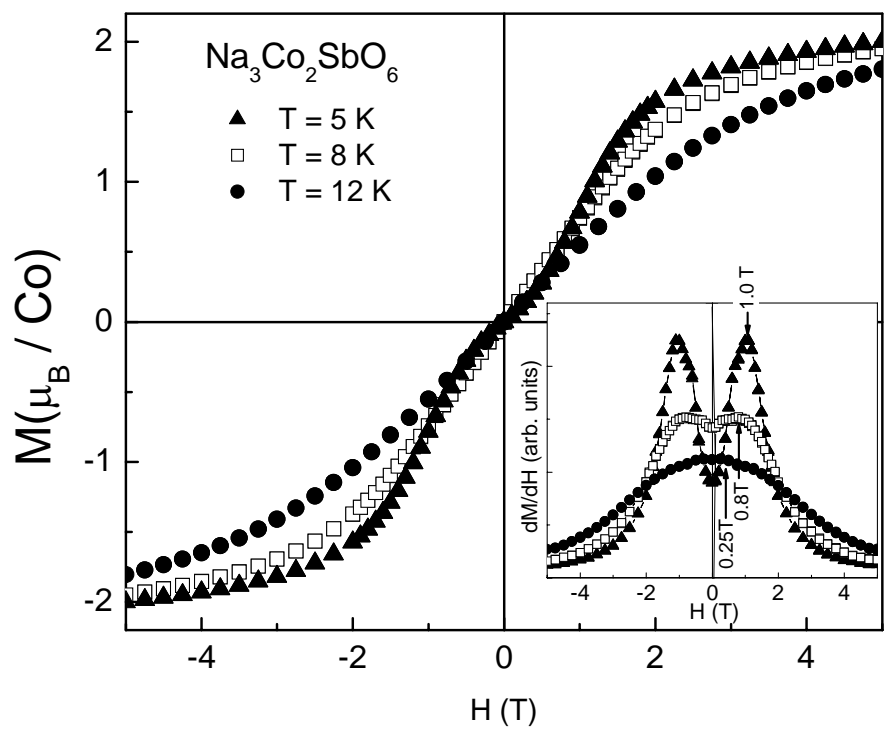

**Figure 8**

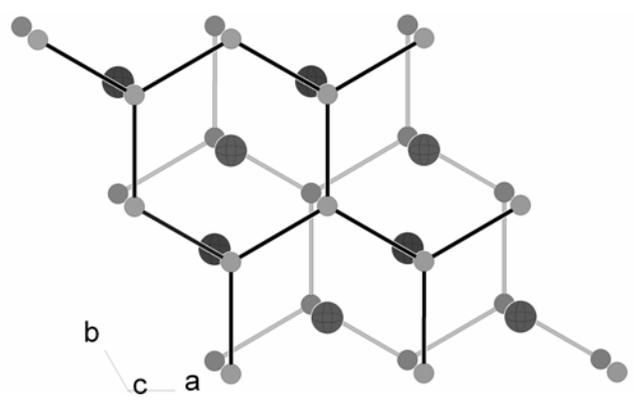

**(a) Na₂Co₂TeO₆**

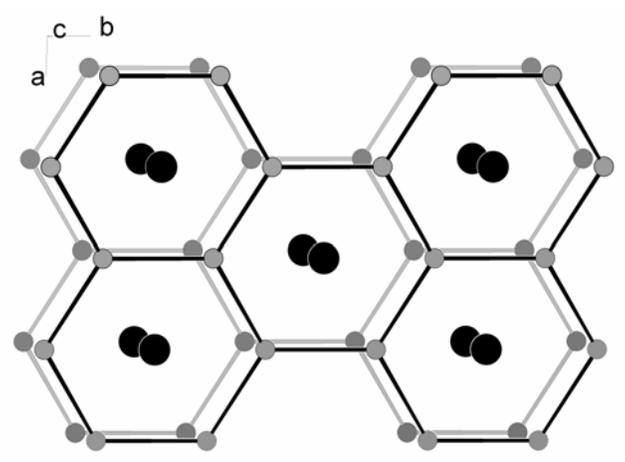

**(b) Na₃Co₂SbO₆**

**Figure 9**